# Growth of honeycomb-symmetrical Mn nanodots arrays on Si(111)-7×7 surface

De-yong Wang, Hong-ye Wu, Li-jun Chen, Wei He, Qing-feng Zhan, and Zhao-hua Cheng*

*State Key Laboratory of Magnetism and Beijing National Laboratory for Condensed Mater Physics, Institute of Physics, Chinese Academy of Sciences, Beijing 100080, China*

## Abstract

The growth of well-ordered Mn nanodots arrays on Si(111)-7×7 reconstructed surface was investigated by means of scanning tunneling microscopy (STM) as well as Kinetic Monte Carlo (KMC) simulation. Mn atoms deposited slowly onto elevated substrates were observed to occupy preferentially on the faulted half unit cells (FHUCs) of Si(111)-7×7 surface. The preference occupancy in the FHUCs, $P_F$, defined as the ratio of number of FHUCs occupied by Mn nanodots to number of all occupied in two halves, decreases with increasing deposition rate as well as decreasing substrate temperature. The KMC simulations, which are in good agreement with the experimental results, were employed to optimize the growth conditions, including deposition rate and substrate temperature, for the self-organized growth of Mn nanodots arrays on Si(111)-7×7 reconstructed surface. By adjusting the deposition rate, one can control the growth of well-ordered and uniform Mn nanodots arrays to form either a triangular symmetry or a honeycomb one.

PACS numbers: 81.15.Aa, 68.35.Bs, 68.35.Fx

*Corresponding author
E-mail:zhcheng@aphy.iphy.ac.cn

## I. Introduction

It is well known that the nanoscaled materials can exhibit significantly different physical and chemical properties from those of corresponding bulk materials due to the quantum size effects. Some properties of the nanodots can alter drastically even when the cluster size varies by a single atom, and the inhomogeneity in an ensemble of clusters will smear their special properties [1]. Therefore, artificially controllable fabrication of nanostructures on supporting substrates with atomic precision is the ultimate goal in fundamental and applied science research [2-6]. Recently, identical Indium, Gallium, Aluminum and Cobalt nanodots have been fabricated self-assembly on Si(111)-7×7 reconstructed surface [7-10]. Si(111)-7×7 reconstructed surface consists of two different types triangular half-unit cells, faulted half-unit cells (FHUCs) and unfaulted half-unit cells (UFHUCs), which are divided by dimmer rows and corner holes. Metal nanodots can occupy either the two halves of unit cells (HUCs) of Si(111)-7×7 reconstructed surface equally to form a honeycomb symmetry or the FHUCs only with a triangular symmetry. In our previous work, well-ordered and uniform Mn nanodots were fabricated on Si(111)-7×7 surface [11]. Due to the relatively large energy difference between adsorption on the UFHUCs and the FHUCs, Mn nanodots deposited onto the elevated substrates were observed to occupy preferentially on the FHUCs of Si(111)-7×7 surface to form a triangular symmetry and honeycomb structure was not formed even at high coverages. In order to increase the density of Mn nanodots and modify the distance of interdots, which may be helpful for investigating the dot-dot interaction and useful for potential application as ultrahigh density recording, we present the controllable growth of uniform Mn nanodots arrays on Si(111)-7×7 surface from triangular symmetry to honeycomb one. Furthermore, the Kinetic Monte Carlo (KMC) simulation was employed to optimize the self-organized growth condition for Mn nanodots arrays on Si(111)-7×7 reconstructed surface.

## II. Experimental details

Our experiments were performed with a combined molecular beam epitaxy (MBE)/scanning tunneling microscope (STM) system in ultrahigh vacuum (UHV) at a base pressure of about $1\times10^{-10}$ mbar. A Si(111) substrate was cleaned by resistive flashing in UHV until a high quality 7×7 reconstruction was observed by OMICRON variable temperature STM. High purity Mn (purity 99.99%) was heated at temperatures ranging from 923 K to 1053 K by a boron nitride crucible, and then deposited onto the surface with deposition rates ranging from 0.17 monolayer (ML)/min(1ML=$7.88\times10^{14}$ atoms/cm$^2$) to 13.4ML/min. All room-temperature scanning tunneling microscopy (STM) images reported here were recorded with tunneling current ranging form 200 to 400 pA. A chemically etched tungsten tip was used as the STM probe.

## III. Results and Conclusions

Since the FHUCs of the Si(111)-7×7 reconstructed surface are more reactive than the UFHUCs and the intrinsic attractive potential wells on the FHUCs effectively trap diffusing metal atoms [12], Mn atoms are expected to occupy preferentially on the FHUCs. In order to control Mn atoms occupy equally on the FHUCs and UFHUCs and form honeycomb structure, one must restrict them to diffuse into the FHUCs. The atomic diffusion can be significantly reduced when they are deposited onto a cooled substrate. Fig. 1(a) is the STM images of Mn nanodots deposited at 145K. Although Mn nanodots distribute randomly on both the FHUCs and UFHUCs of Si(111)-7×7 surface, the diameter and height distributions are very broad(Figs. 1(b) and 1(c)). The dispersions, $\Delta d = (<d^2> - <d>^2)^{1/2}$ and $\Delta h = (<h^2> - <h>^2)^{1/2}$ are 0.31 nm and 0.028 nm, respectively, which are about 23.9% and 12.4% of the average diameter and height values. The broad size distribution of Mn nanodots is undesirable for future applications.

The size distribution of Mn nanodots can be significantly narrowed when the substrate temperature is heated up to 450 K. Fig. 2(a), (b), (c) show the STM images of Mn nanodots deposited onto the Si(111) substrate heated to 450 K with three different deposition rates of 0.17 ML/min., 0.64 ML/min, and 1.8 ML/min., respectively. The diameter and height histograms of the nanodots are illustrated in Figs.3(d)-(i). The dispersions, $\Delta d$ and $\Delta h$ are found to decrease to about 10% of the average diameter and height values when the substrate temperature increases up to 450 K. With increasing deposition rate, the size and shape of Mn nanodots remain essentially unchanged. At elevated substrate temperature, the deviation in size distribution can be eliminated by atom hopping quickly in half unit cells (HUCs). In the case of Mn nanodots on Si(111)-7×7 surfaces, the attractive potential wells formed by the triangular HUCs(2.7 nm) restrict Mn nanodots to overgrow the HUCs boundaries at low coverage, and consequently result in the narrow size distribution.

Fig. 3(a) illustrates the preference in the FHUCs occupancy, $P_F$, (ratio of number of FHUCs occupied by Mn nanodots to number of all occupied in two halves) as a function of coverage. Mn atoms occupy the FHUCs, which is energetically the most favorable in the case of low coverage. It is surprising that $P_F$ is almost invariable with increasing coverage even up to 0.22 ML. With further increasing the coverage, Mn nanodots will aggregate into big islands across the HUCs boundaries. Since the preference $P_F$ is kinetically determined by the ratio between hopping rate and deposition rate, maximum value of $P_F$ can be achieved when Mn adatoms have enough time for rearranging on the surface during growth. When the deposition rate is too high, they form sizeable clusters prior to diffusing into the more stable position. Therefore, $P_F$ is expected to decrease with increasing the deposition rate, as plotted in Fig. 3(b).

Detailed behaviors for atoms on surface are highly desirable to understand the underlying formation mechanism of these nanodots. However, the *in-situ* observation of the atoms behaviors is very difficult at currently experimental conditions. The deeper understanding about the kinetics of the dots growth on Si(111) surface has been obtained with the help of Kinetic Monte Carlo simulation (KMC simulation). The coarse grained KMC model implements were described in details elsewhere [13, 14]. Mn atoms are assumed to fall into initially the two half-unit cells(HUCs) randomly. Once an atom falls into a HUC, it immediately moves to its adjacent occupied HUCs if they exist. In this model, isolated cluster cannot overgrow HUC boundaries, the capacity of a HUC to accommodate atoms can not exceed $n_s$. When the number of atoms in a cluster $n \leq n^*$ ($n^*$-critical size), the cluster can decay by hopping out atoms one by one. The activation energy for one atoms hopping out of a HUC containing $n$ atoms is given by $E_n = E_{F,U} + (n-1)E_a$, $E_{F,U}$ is the barrier for single Mn atom hopping out of FHUCs and UFHUCs, respectively, $E_a$ is the effective binding energy between adatoms inside a HUCs. The attempt frequency $v_0$ is assumed to be the same for the FHUCs and UFHUCs, and attempts of Mn atoms to hop out of a HUC are considered to be independent, so that the hopping rate of a Mn atom can be written as $v_n^{F,U} = nv_0^{F,U} \exp(-E_n/kT)$. Our simulation consists of growth in 450K and a subsequent annealing process at room temperature for an hour. By fixing $v_0^{F,U} = 5 \times 10^9 /s$, $n^*$=5, $n_s$=21[14] and changing $E_F$, $E_U$, $E_a$, randomly, we found a least square between simulation and experimental data points. The simulated $P_F$ as a function of deposition rate is shown in Fig. 3(b). The model parameters which provide the best agreement with experimental data is: $E_F$=0.72 *eV*, $E_U$=0.70 *eV*, $E_a$=0.05 *eV*. The (observed) qualitative agreement suggests that the model correctly captures the basic features of the growth process for Mn nanodots on Si(111)-7×7 surface. By fixing the deposition rate to 0.17ML/min., the effect of substrate temperature on the preference $P_F$ was also observed by STM and KMC simulation. As shown in Fig.3(c), both KMC simulation and STM images observation demonstrate consistently that a weak preference of Mn nanodots in the FHUCs can be obtained when the substrate is cooled to low temperatures. Therefore, the KMC simulations were employed to optimize the growth conditions, including deposition rates and substrate temperatures, for the self-organized growth of Mn nanodots arrays on Si(111)-7×7 reconstructed surface, which is illustrated in Fig. 4. It reveals that the maximum value of preference in the FHUCs occupancy, $P_F$, can be achieved when Mn deposited onto an elevated substrate with very low deposition rate. The preference can be weakened either by decreasing substrate temperature or by increasing the deposition rate and keeping the substrate temperature at a certain temperature.

Figs. 5(a) and 5(b) illustrate two typical STM images of Mn nanodots with deposition rates of 0.17 ML/min. and 13.4 ML/min., respectively. A triangular symmetry of Mn nanodots array is observed when the deposition rate is very low. With increasing the deposition rate up to 13.4 ML/min., the triangular symmetry of Mn nanodots transforms to honeycomb structure due to Mn occupy equally two halves of unit cells of Si(111)-7×7 reconstructed surface. Therefore, the controllable growth of well-ordered and uniform Mn Nanodots arrays with a honeycomb symmetry can be achieved on Si(111)-7×7 surface.

### IV. Summay

In summary, both STM observation and KMC simulation indicate that the reduction of preference in the FHUCs occupancy, $P_F$, can be achieved either by decreasing the substrate temperature or by increasing the deposition rate and keeping the substrate temperature at a certain temperature. However, the diameter and height of Mn nanodots deposited onto a cooled substrate show a very broad distribution. Due to the attractive potential wells formed by the triangular HUCs(2.7 nm), which restrict to overgrow at HUCs boundaries, uniform Mn nanodots arrays can be ordered a honeycomb structure on Si(111)-7×7 surface with substrate temperature of 450 K by increasing the deposition rates.

**Acknowldgements**

This work was supported by the State Key Project of Fundamental Research, and the National Natural Sciences Foundation of China.

**References**

[1]W. P. Halperin, Rev. Mod. Phys. **58**, 533 (1986).

[2] P. Moriarty, Rep. Prog. Phys. **64**, 297 (2001).

[3] V. A. Shchukin and D. Bimberg, Rev. Mod. Phys. **71**, 1125(1999).

[4]. P. Gambardella, S. Rusponi, M. Veronese, S.S. Dhesi, C.Grazioli, A. Dallmeyer,, I. Cabria, R. Zeller, P.H. Dederichs, K. Kern, C. Carbone, and H. Brune, Science, **300**,1130(2003).

[5]S. Morup and C. Frandsen, Phys. Rev. Lett. 92,217201 (2004).

[6] C. Demangeat and J. C. Parlebas, Rep. Prog. Phys. **65,** 1679 (2002)

[7] J.L. Li, J.F. Jia, X.J. Liang, X. Liu, J.Z. Wang, Q.K. Xue, Z.Q. Li, J.S. Tse, Z.Y. Zhang and S.B. Zhang, Phys. Rev. Lett. **88, 066101** (2002).

[8] M. Y. Lai and Y. L. Wang, Phys. Rev. B **64**, 241404 (2001)

[9]J.F. Jia, X. Liu, J.Z. Wang, J.L. Li, X.S. Wang, Q.K. Xue, Z.Q. Li, Z.Y. Zhang and S.B. Zhang, Phys. Rev. B. 66, 165412(2002)

[10]C.Pirri, J.C. Peruchetti, G. Gewinner and J. Derrien, Phys. Rev. B 29 3391 (1984); P.A. Bennett, D.G. Cahill and M. Copel, Phys. Rev. Lett. **73**, 452(1994).

[11]D.Y. Wang, L.J. Chen, W. He, Q.F. Zhan and Z.H. Cheng, J. Phys. D: Appl. Phys. **39**(2006)(in press)

[12] K. Takayanagi, Y.Tanishiro, M.Takahashi, and S. Takahashi, J. Vac. Sci. Technol. A 3,1502(1985)

[13] J. Myslivecek, P. Sobotık, I. Ostadal, T. Jarolımek, P.Smilauer, Phys. Rev. B 63 (2001) 45403.

[14] P. Kocan, P. SobltiK, I. Ostadal, and M. Kotrla, Surf. Sci. 566-568 (2004) 216.

**FIGURE CAPTIONS**

Fig. 1(color online): STM images of Mn dots deposited at temperatures of about 145K with the coverage of 0.21 ML (a) and the corresponding histograms of diameter (b) and height(c). The STM image was obtained at Vs= -2.093V and scanned area of $30\times30$ nm$^2$.

FIG. 2(color online) (a), (b), and (c) are STM images of Mn nanoclusters deposited on the Si(111)-7×7 at about 450K with different Mn fluxes. Images were obtained with the sample bias of -2.2V and tunneling current of 0.25nA. Below the STM images are the corresponding histograms of cluster diameter and height, (d)-(i), respectively.

Fig.3. (color online) Preference factor $P_F$ as a function of coverage for Mn nanodots deposited onto the substrate with temperature of 450 K (a), deposition rates(b), and substrate temperatures(c).

Fig.4. (color online) Simulated diagram of growth conditions, including deposition rates and substrate temperatures, for the self-organized growth of Mn nanodots arrays on Si(111)-7×7 reconstructed surface.

Fig. 5. (color online) Typical STM images of Mn nanodots with deposition rates of 0.17 ML/min. (a) and 13.4 ML/min.(b). Images were obtained with the sample bias of -2.5V and tunneling current of 0.25 nA.

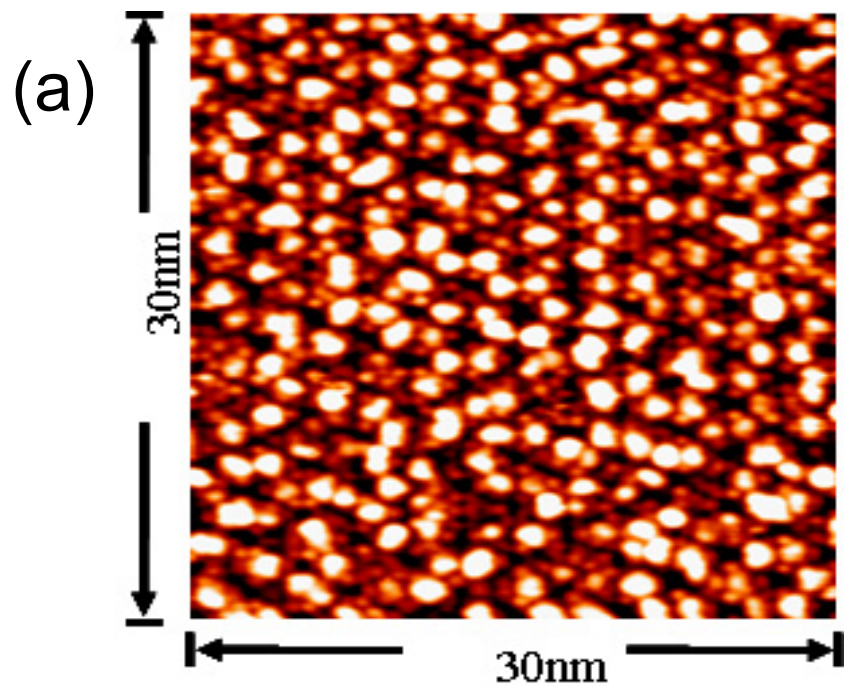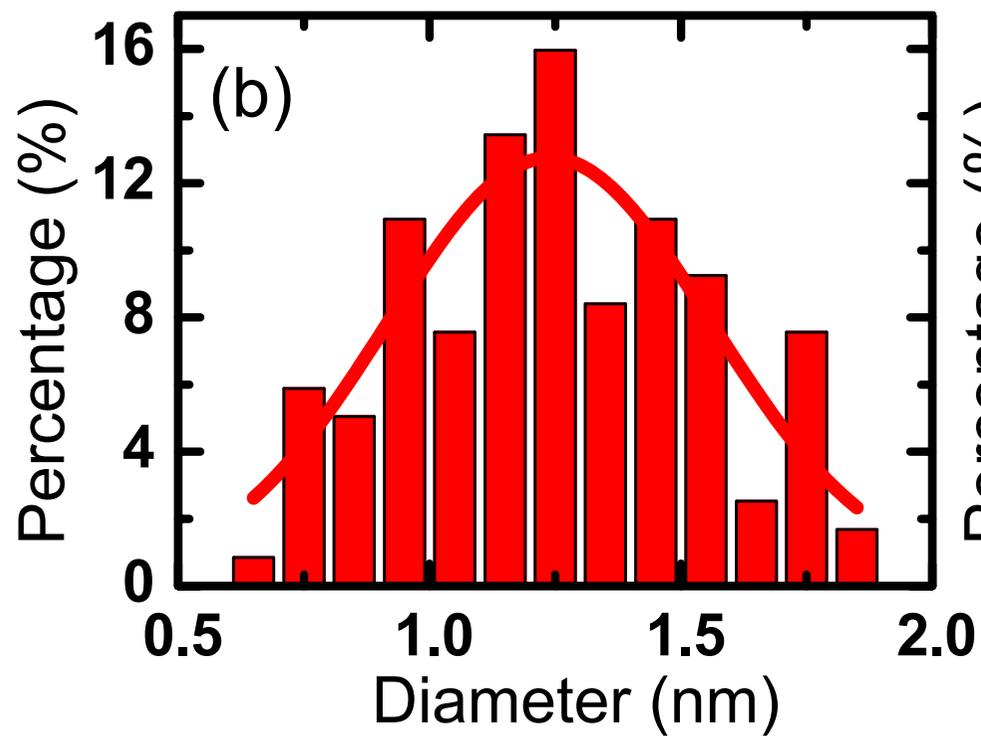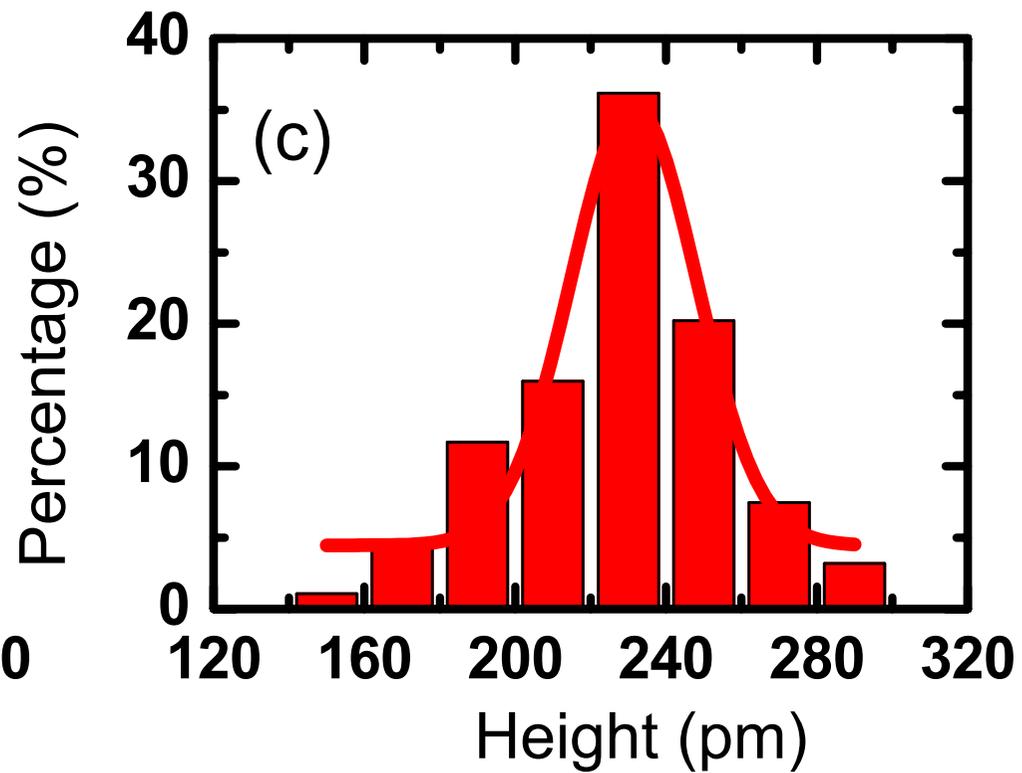

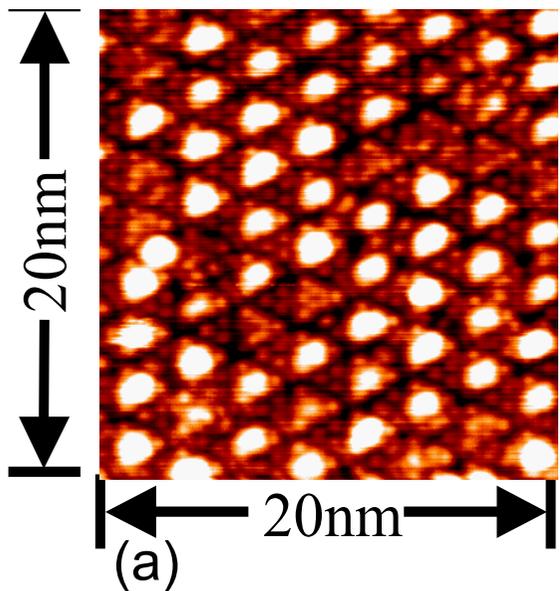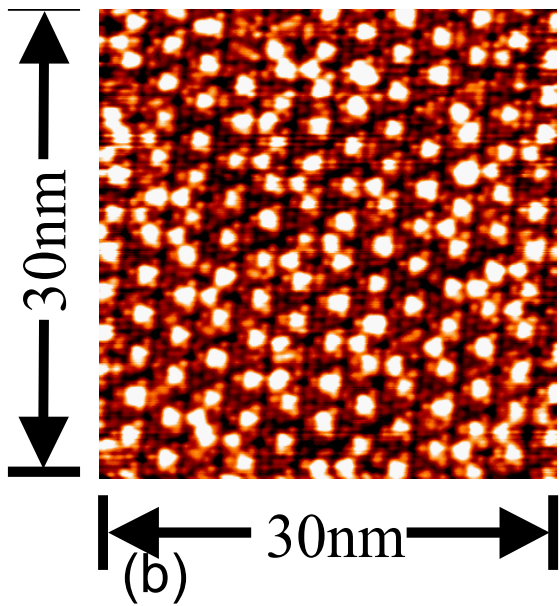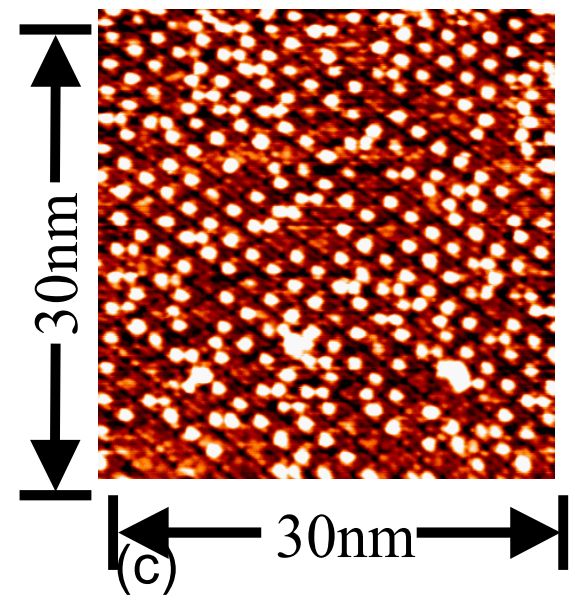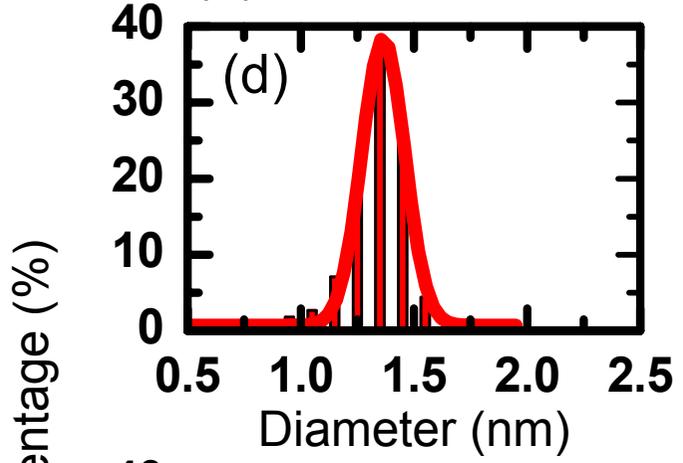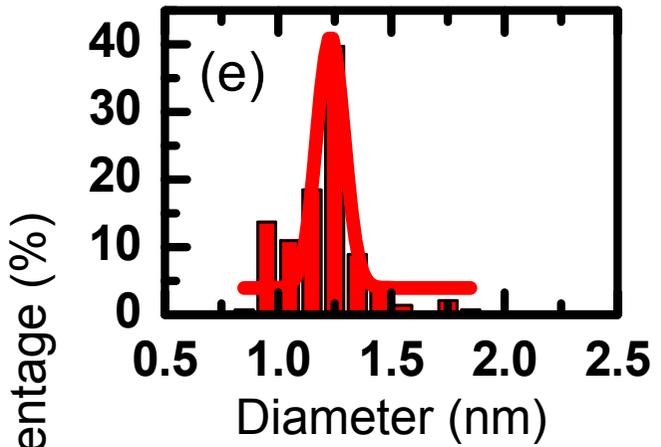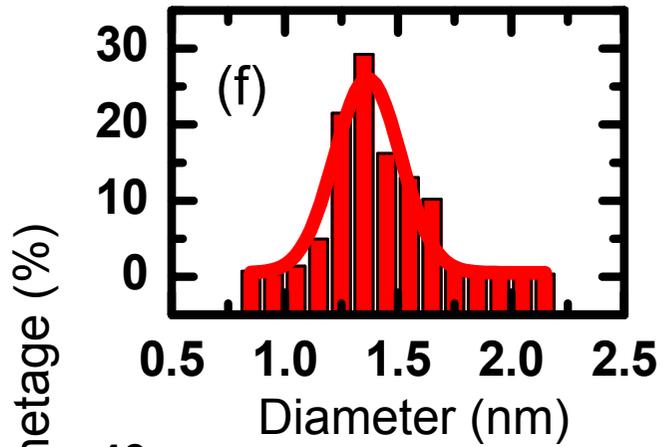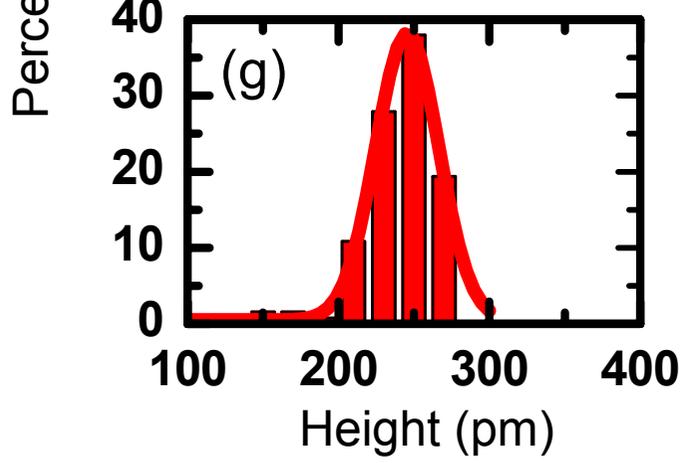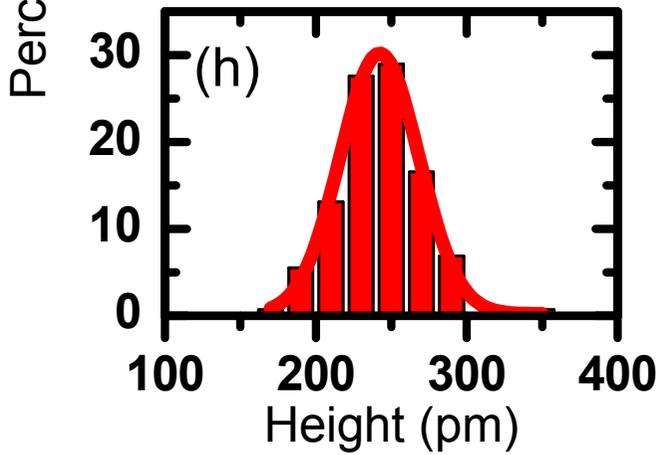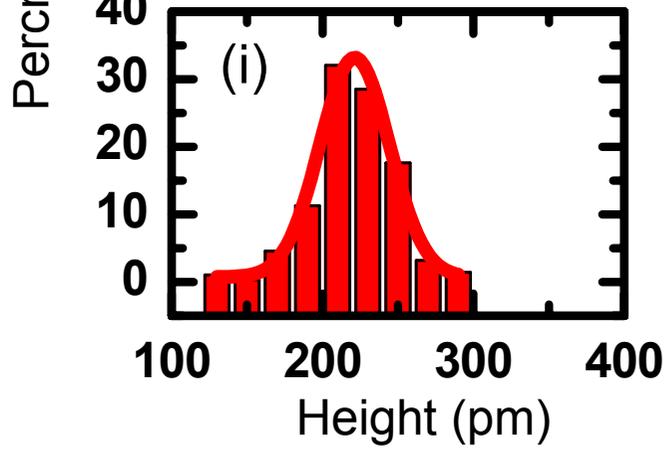

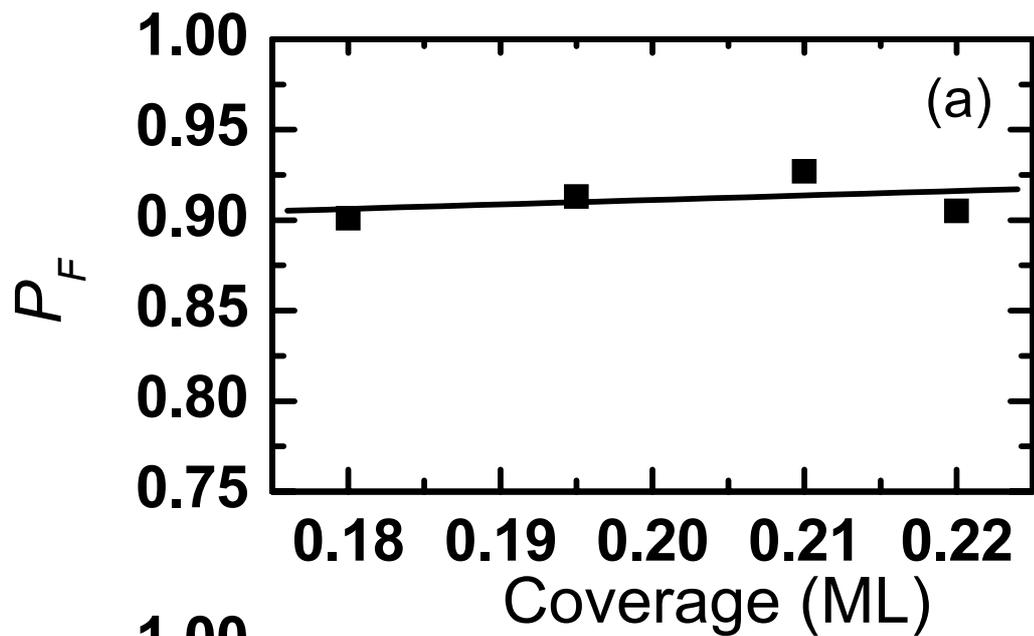

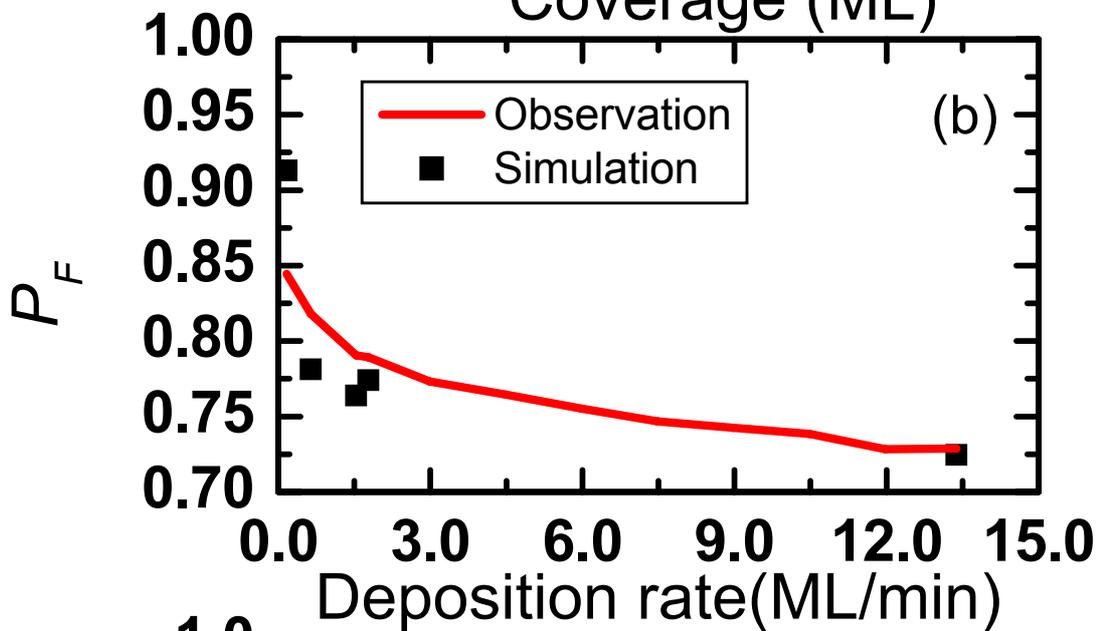

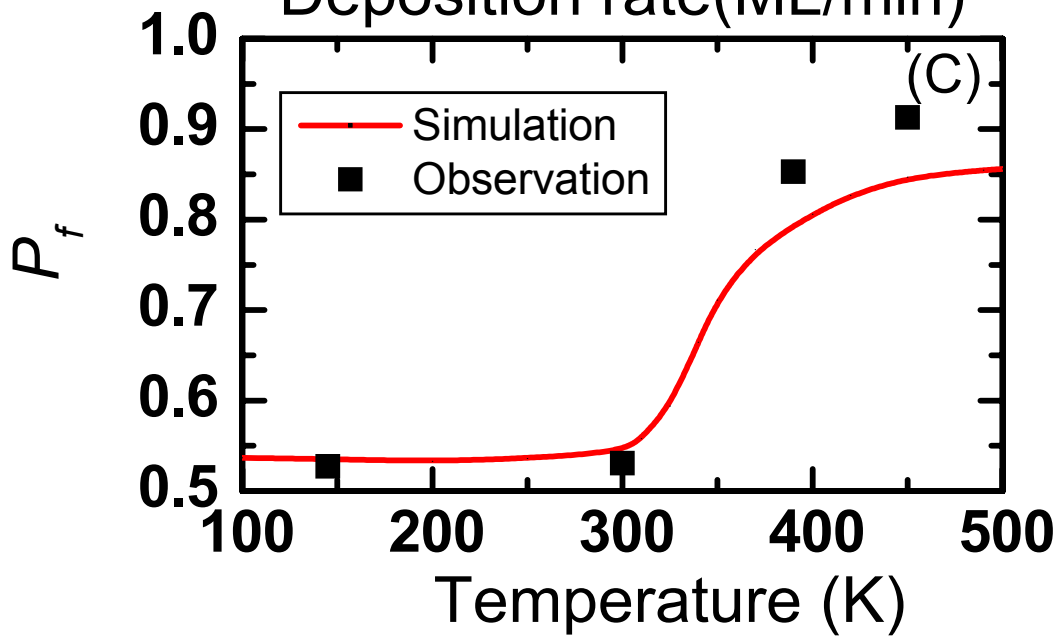

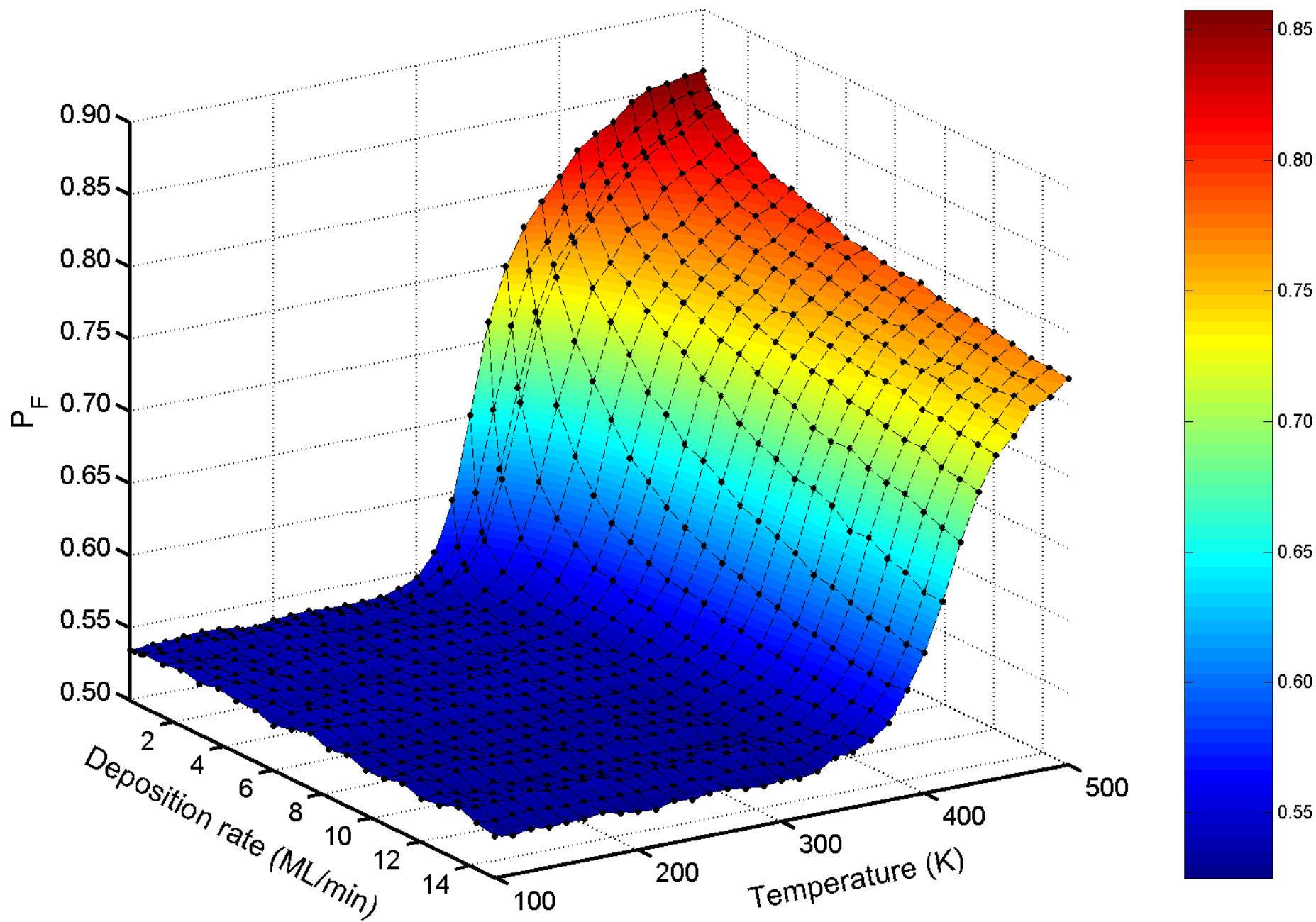

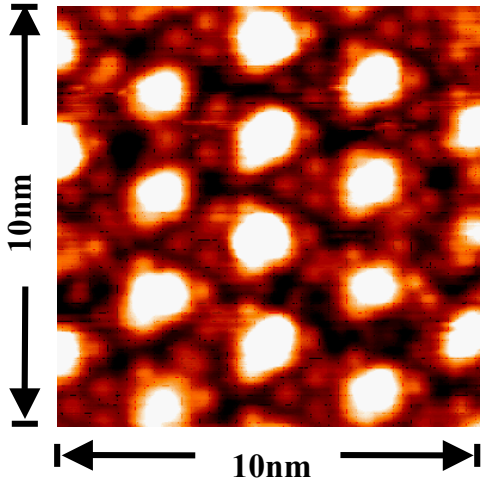 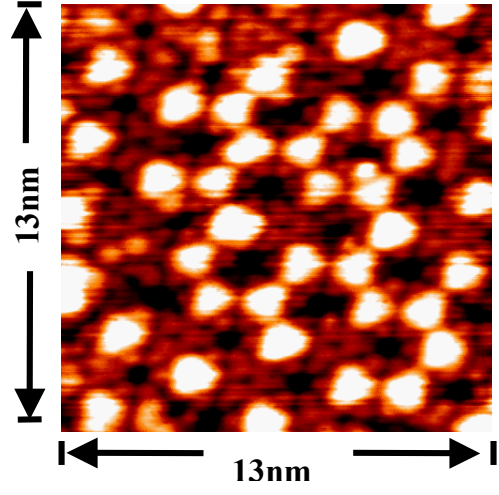